\documentclass{JHEP3} 

\usepackage{epsfig,multicol,bbm,amsmath}
\usepackage{graphicx}
\usepackage{amssymb}
\usepackage{mathrsfs}

\newcommand\beq{\begin{equation}}
\newcommand\eeq{\end{equation}}
\newcommand{\bea}{\begin{eqnarray}}
\newcommand{\eea}{\end{eqnarray}\noindent}

\graphicspath{{fig/}}

\title{Chiral Modulations in Curved Space I: Formalism}

\author{Antonino Flachi \\
        Yukawa Institute for Theoretical Physics and Department of Physics, Kyoto University, Kyoto, Japan\\
        E-mail: \email{flachi@yukawa.kyoto-u.ac.jp}}

\author{Takahiro Tanaka \\
        Yukawa Institute for Theoretical Physics, Kyoto University, Kyoto, Japan\\
        E-mail: \email{tanaka@yukawa.kyoto-u.ac.jp}}

\received{} 		
\accepted{}		

\preprint{KUNS-2313; YITP-10-102}		

\abstract{The goal of this paper is to present a formalism that allows to handle
 four-fermion effective theories at finite temperature and density in
 curved space. The formalism is based on the use of the effective action
 and zeta function regularization and supports the inclusion of
 inhomogeneous and anisotropic  phases.
One of the key points of the method is the use of
 a non-perturbative ansatz for the heat-kernel that returns the
 effective action in partially resummed form, providing a way to go
 beyond the approximations based on the Ginzburg-Landau expansion for
 the partition function. The effective action for the case of
 ultra-static Riemannian spacetimes with compact spatial section is
 discussed in general and a series representation, valid when the
 chemical potential satisfies a certain constraint, is derived. To see
 the formalism at work, we consider the case of static Einstein spaces at
 zero chemical potential. Although in this case we expect inhomogeneous
 phases to occur only as meta-stable states, the problem is complex enough
 and allows to illustrate how to implement numerical studies of
 inhomogeneous phases in curved space. Finally, we extend the formalism
 to include arbitrary chemical potentials and obtain the analytical
 continuation of the effective action in curved space.
}

\keywords{quantum fields in curved space; chiral fermions; zeta function regularization}

\begin{document}

\section{Introduction}

{
The Nambu-Jona Lasinio (NJL) \cite{nambu} and the Gross-Neveu (GN)
\cite{gross} models are the two most notable examples of four-fermion
effective theories ($4f$ET) sharing the global symmetries of QCD as well
as displaying the phenomenon of chiral symmetry breaking. $4f$ET have
been attracting attention since their inception and it is now well
understood that they provide useful working models allowing to describe
dynamical chiral symmetry breaking in vacuum and in hot-dense baryonic
matter, to investigate the QCD phase diagram and, in general, low-energy
non-perturbative effects involving strong interactions at finite
temperature and density (There are many reviews available in the
literature, we mainly consulted
Refs.~\cite{review1,review2,review3,review4,inagaki,kunihiro}). 

$4f$ET, aside from being central in the study of superconductivity, quark condensation, the physics of light mesons, just to mention a few, cover remarkable importance in describing many astrophysical and cosmological systems where the effects of strong interactions play a prominent role. Neutron stars and astrophysical compact objects are, in fact, a natural playground for such models. Interesting ramifications also exist in connection with the hadrosynthesys in the early universe.
Importantly, due to the experimental efforts currently carried out at
heavy ion colliders and directed to explore the properties of QCD at
high temperature and density, the above problems acquire a timely
importance also from a phenomenological point of view.

$4f$ET are usually discussed within the mean-field and large-$N$ approximations, with the condensate defining the different phases assumed to be spatially homogeneous. However, there are various reasons to believe that there may be inhomogeneous phases in the phase diagram of strongly interacting theories as well. Recent attention has been focusing on trying to understand whether these phases may actually form. Although the study of these phases is very important from both a theoretical and a phenomenological perspective, it is so far rather limited, mostly due to technical complexity. Amongst the cases discussed so far, the formation of inhomogeneous condensates in NJL-class of models has received some attention. An initial analysis has been performed within the chiral density wave approach in which the spatially varying order parameter is assumed to be a plane wave \cite{nakano}. In Refs.~\cite{nickel1,nickel2} a Ginzburg-Landau approximation complemented by a full numerical study has been used. The results indicate that, in the vicinity of the critical chiral point, the phase diagram is carachterized by two second order phase transitions tracks, intersecting at a Lifshitz point, with an inhomogeneous ground state rensembling a lattice of domain walls, which seems to be favoured in the vicinity of the chiral critical point. Solitonic ground states in color superconductivity have also been analyzed in Ref.~\cite{alford,bowers,mannarelli,rajagopal,buballa}. Other relevant analyses were also performed in the context of the GN model in $1+1$ dimensions that allows for exact solutions (see, for instance, Refs.~\cite{basar1,basar2,basar3,schnetz,urlichs}).

We are especially interested in understanding how the effect of an external, spatially varying gravitational field may modify the phase diagram of QCD. This is clearly a very complicated problem and the aim of this paper is to set up a formalism useful to study the above issue. Our goal is to develop a different approach, based on the use of the effective action formalism and zeta function regularization, which allows for generalization to curved space and the inclusion of inhomogeneous and anisotropic phases. As we will see, the method described in this paper may also be easily adapted to include varying external fields of various sorts, whose analytical study is so far limited to the case of spatially homogeneous, and/or weak fields. 

Understanding how the phase structure of $4f$ET may be modified due to the presence of external gravitational fields has received some attention (see, for example, Refs.~\cite{eb2,huang}). The approach tipically used for accounting the presence of the gravitational field is based on the expansion of the fermion propagator in powers of the external field and is limited to the case of homogenous condensates, which is usually not reliable in the vicinity of the transition. A better approach based on the direct computation of the effective potential without relying on a weak curvature expansion can be easily designed as long as the condensate, curvature and any additional external field are homogeneous. Such a direct approach has been used in Ref.~\cite{huang} to obtain the phase structure of the NJL model on a cosmological background.

Developing an efficient method that allows the inclusion of inhomogeneous condensates is one of the aims of this work. In flat space this problem has been considered in Ref.~\cite{nickel2}, which uses the Ginzburg-Landau (GL) expansion for the partition function. The GL-expansion takes the following form:
\bea
\mathscr{S}_{GL} &=& {\alpha_2 \over 2} \sigma^2 + {\alpha_4 \over 4} \left[\sigma^4 + \left(\nabla \sigma\right)^2\right] + {\alpha_6 \over 6}\left[ \sigma^6 + 5 \left(\nabla \sigma\right)^2 \sigma^2 +{1\over 2} \left(\Delta \sigma\right)^2 \right]+\cdots~,
\label{gle}
\eea
where the dots stand for higher order terms, and has the limitation of leaving out non-perturbative effects. 
The above expression is general and, once the underlying model is fixed, the coefficients
$\alpha_n$ can be computed. 

As an alternative, in principle, one can design a method based on the direct computation of the effective action, in a similar way to what is done in Ref.~\cite{huang} that considers constant chiral condensates on the background of a static Einstein space. The method used there is, however, based on the explicit knowledge of the eigenvalues of the Dirac operator. For this reason this approach simply cannot be adapted to include the general cases with spatially varying condensates. 

In the following we will present a method that bypasses the drawbacks of the above approaches and allows to compute the effective Lagrangian on a  curved background of ultrastatic type (or conformally related to an ultrastatic manifold). 
When we consider spatially inhomogeneous background space-times (or external fields), 
we also need to care about the regularization procedure. To avoid ambiguity related to regularization, we adopt a method
that keeps the general covariance manifest.  

In the next section we will briefly remind the basics, and in Sec.~\ref{sec3} we will illustrate our formalism. 
The method allows to compute 
the effective Lagrangian for the condensate 
in partially resummed form, which can be used as a starting point for
analyzing the properties of the inhomogeneous phases and the structure
of the phase diagram. This, in general, requires a non-trivial numerical
effort. In order to see the method at work and to anticipate possible
complexties in the numerical analysis, we consider the case of
inhomogeneous condensates in a static Einstein space at zero chemical
potential. This case is still complicated enough to deserve discussion and
will help us to understand better what are the possible issues in
applying the method to cases of physical interest. This example will be
discussed at length in Sec.~\ref{sec4}, where we develop the necessary
numerics to investigate whether inhomogeneous phases may form due to
non-trivial curvature, finite size, and non-perturbative effects. In fact, we will find that inhomogenous kink-type
solutions form. These solutions, however, have free
energy larger than the corresponding homogeneous phases, and therefore
they can only appear as excited states.
Sec.~\ref{sec5} is devoted to extend the formalism to aritrary chemical potentials and conclude in Sec.~\ref{sec6}. Some technical details are included in Appendix.}

\section{Four-fermion Effective Theories}
\label{sec2}

In the following we will consider as a prototype model of $4f$ET the following $D$-dimensional theory:
\bea
S= \int d^Dx \sqrt{g} \left\{ 
\bar \psi i \gamma^\mu \nabla_\mu \psi 
+ 
{G\over 2N} \left(\bar \psi \psi\right)^2 + \cdots
\right\}~,
\label{action}
\eea
where $\psi$ is a $(D \times N_f\times N_c)$-component quark spinor,
with $N_f$ flavors and $N_c$ colors ($N\equiv N_f \times N_c$),
$\gamma_\mu$ are the gamma matrices in curved space, $G$ is the
four-fermion coupling constant and $g=|\mbox{Det} g_{\mu\nu}|$. The rest
of the notation is standard \cite{parkertoms,odints}. The dots stand for terms with higher mass dimension.

Finite temperature
will be introduced by means of the imaginary time formalism and finite
density by means of a chemical potential contribution of the form $\mu
\bar\psi \gamma^0 \psi$. In the following, we will not consider how the
chiral anomaly may be induced by gravitational effects, and thus we neglect
terms of the form $\left(\bar\psi \gamma^5 \psi\right)^2$. 

The above action is invariant under discrete chiral transformations and mass terms cannot appear without breaking chiral symmetry. On the other hand, if the chiral symmetry is broken dynamically, the composite operator $\langle \bar \psi \psi \rangle$ acquires a non-zero vacuum expectation value and a fermion mass term would appear. The action is also invariant under SU($N_f$) flavor symmetry.
In the following, we will stick to the large-$N$ approximation. 

Starting from the theory in flat space allows for a
direct generalization to curved space by replacing the flat space metric
and ordinary derivatives with the metric tensor and covariant
derivatives. This process of `covariantization' also requires to augment
the original action with all the terms compatible with coordinate
invariance and the symmetries of the action. 
The leading contribution is $R (\bar \psi \psi)^2$, and it is suppressed relative to 
the four-fermion interaction term by the ratio between the curvature and the fundamental 
energy scale squared.  
Higher order terms are further suppressed by inverse powers of the fundamental energy scale and/or inverse powers of $N$.
In the four dimensional case, for example, 
the divergences in the effective action contain a term proportional to $R (\bar \psi \psi)^2$, 
but not any subleading term (in the sense defined above). 
In this case, 
after calculation of loop corrections, one has to determine 
the renormalized value of the coefficient of $R (\bar \psi \psi)^2$ 
according to an experimental input. However, such an experimental 
input is lacking for this term. 
Therefore, one renormalization parameter is inevitably left unfixed. 
As we can adjust the renormalized coefficient 
of $R (\bar \psi \psi)^2$ later 
by means of finite renormalization, 
in the following calculation we choose the zeroth order 
Lagrangian not to contain the term $\propto R (\bar \psi \psi)^2$, 
which does not lose generality as long as only the leading correction 
due to curvature $R$ is concerned.

The basic formalism will be illustrated for the case of a $D=d+1$ dimensional, ultra-static spacetime of the form 
\bea
ds^2 = dt^2 - g_{ij} dx^i dx^j ~,
\eea
where the tensor $g_{ij}$ represents the metric on the spatial section $\mathscr{M}$ of the spacetime. We stress, however, that the procedure described below is always possible, with minor modifications, if the spacetime can be conformally related to an ultra-static one. 

Allowing a mean field value $\langle \bar\psi \psi \rangle = -N\sigma(x)/G$ for the chiral condensate, after bosonization, the partition function can be expressed as a path integral over $\sigma$:
\beq
Z = \int \left[d\sigma\right] e^{i\mathscr{S}_{eff}}~,
\eeq
where the effective action (per fermionic degree of freedom) at the lowest order in the large-$N$ approximation can be written as
\bea
\mathscr{S}_{eff} &=& 
-\int d^{D}x \sqrt{g} \left({\sigma^2\over 2G}\right) 
+
\ln \mbox{Det} \left( i \gamma^\mu \nabla_\mu - \sigma + \mu \gamma^0\right)~.
\eea
The above determinant acts on the Dirac spinor, and coordinate
space. The last term, proportional to $\mu$, represents a chemical
potential term that allows to include finite density effects, while
finite temperature is introduced using the imaginary time formalism,
$t\rightarrow -i\tau$ with period $\beta = 2\pi/T$, and imposing
anti-periodic boundary conditions on the fermion fields,
$\psi(\tau)=-\psi(\tau+\beta)$. 

The effective action can be expressed as \cite{parkertoms}
\bea
\mathscr{S}_{eff} &=& -\int d^{D}x \sqrt{g} \left({\sigma^2\over 2G}\right) +{1\over 2} \ln \mbox{Det} \left[
\Box +{1\over 4}R +\sigma^2 -\mu^2 -2 i \mu {\partial\over \partial t} + i \gamma^\mu \left(\nabla_\mu \sigma\right)
\right]
~.\label{dseft}
\eea
Finally, making finite temperature effects explicit leads to
\bea
\mathscr{S}_{eff} = -\int d^{D}x \sqrt{g} \left({\sigma^2\over 2G}\right) 
+{1\over 2} \sum_\lambda \sum_{n=-\infty}^\infty 
\ln \mbox{Det}\, \mathscr{D}^{(n)} 
~,
\label{las}
\eea
where we defined $d$-dimensional operators 
\bea
\mathscr{D}^{(n)} 
\equiv - \Delta + \omega_n^2 + {1\over 4}R +\sigma^2 -\mu^2 -2 i \mu \omega_n  + \lambda \left|\partial \sigma\right|~,
\label{27}
\eea
with the frequencies given by
\bea
\omega_n={2\pi\over \beta}\left(n+{1\over 2}\right)~,
\eea
and
\bea
\Delta = {1\over \sqrt{g}} \partial_i \left(\sqrt{g}g^{ij}\partial_j\right)~,
\eea
being the Laplacian over the spatial section $\mathscr{M}$. 

To obtain formula (\ref{las}), we chose one of the directions of 
our tetrad frame to coincide with the direction in which the condensate varies. Namely, 
$e_{(1)}^j:=\partial^j\sigma/|\partial \sigma|$. 
Then, we decomposed the 
$2\left[D/2\right] \times 2\left[D/2\right]$ matrix operator 
in Eq.~(\ref{dseft}) (where a square bracket means the floor function) 
by means of the eigenvectors 
of the first component of the gamma matrix in the tetrad frame, 
$\gamma^{(1)}$, defined by 
(the $-i$ factor and choosing $e_{(1)}^j$ as preferred direction is just due to our convention),
\bea
\gamma^{(1)} \psi_{\lambda} = -i \lambda \psi_{\lambda}~, 
\eea
with $\lambda=\pm 1$. 
In Eq.~(\ref{las}), the summation over $\lambda$ is taken for all 
$2\times[D/2]$ eigenvalues. 
For example, in the case of spherical symmetry, $\sigma=\sigma(r)$, 
identifying $\gamma^r =\sqrt{g^{rr}}\gamma^{(1)}$, 
the last term in (\ref{27}) becomes $\lambda 
\sqrt{g^{rr}}\sigma'(r)$. 

It is immediate to see that if we keep the background space homogeneous and isotropic, as well as the condensate $\sigma$, one can obtain the eigenvalues of the
operator. Then the effective potential can be computed exactely (no
explicit knowledge of the eigenvalues is, in fact, necessary). If we
remove the assumption concerning the background, then such a direct
approach wouldn't work in general. 

Most of the next section will be devoted to describe the details of the computation of the above functional determinant with $\sigma$ assumed to be spatially varying.

\section{Computation of the Effective Action}
\label{sec3}
The effective action can be formally expressed as
\bea
\mathscr{S}_{eff} &=& -\int d^Dx \sqrt{g}  \left({\sigma^2\over
2G}\right) + \delta \Gamma~,
\eea
with
\bea
\delta \Gamma = {1\over 2} \int d^dx \sqrt{g}  \left( \zeta(0) \ln \ell^2 + \zeta'(0)\right)~,
\label{dg}
\eea
where $\zeta(s)$ is the zeta function associated with $\mathscr{D}^{(n)}$ and $\ell$ a renormalization length scale. 
The values of the zeta function and its derivative at $s=0$ are understood as regularized by means of analytical continuation.

In the present case, it is convenient to define the zeta function in terms of the Mellin transform of the heat-trace,
\bea
\zeta (s) =  {1\over \Gamma(s)} \sum_{n, \lambda} \int_0^\infty dt\,
t^{s-1} \mbox{Tr}\,e^{-t \mathscr{D}^{(n)}}~.
\label{zetan}
\eea
A connection with the expanded form for the GL partition
function (\ref{gle}), can be made explicit by using a local expansion
for the heat-trace \cite{gilkey},
\bea
\mbox{Tr}\,e^{-t \mathscr{D}^{(n)}} = {1\over (4\pi t)^{d/2}}
\sum_{j=0}^{\infty} \mathscr{H}_j\, t^j~,
\eea
where the coefficients
$\mathscr{H}_j$ are the standard heat-kernel coefficients now tabulated
in many places (see, for example, Ref.~\cite{vassilevich}). 
The reader may easily notice the analogy between the GL expansion and the above heat-kernel expansion of the functional
determinant in powers of $\sigma$ and its derivatives.
In this sense, the heat-kernel expansion is a generalization of the GL expansion to curved space. 
It may be worth noticing that this approach can also be adapted to the case of
manifolds with boundaries just by modifying the heat-kernel coefficients
that will acquire, aside from global contributions, boundary
terms. 

Conveniently, it is possible to do better by using a
partially resummed form for the heat-trace that would correspond to a
partially resummed form for the GL-expansion. 
A better ansatz for the heat-trace is
\bea
\mbox{Tr}\,e^{-t \mathscr{D}^{(n)}} = {1\over (4\pi t)^{d\over 2}} e^{-t \mathscr{Q}} \sum_k \mathscr{C}^{(k)}_{\lambda}\, t^k~,
\label{ht}
\eea
with $\mathscr{Q}= \omega_n^2 + R/12 +\sigma^2 -\mu^2 -2 i \mu \omega_n  + \lambda \left|\partial \sigma\right|$.
The above form for the heat-kernel in curved spacetime was conjectured in Ref.~\cite{parker} and demonstrated in Ref.~\cite{jack}. The result tells that, for an operator of the form $-\Box +E(x)$ on a Riemannian manifold, all powers of the scalar curvature $R$ and the function $E(x)$ are generated by the overall exponential factor in (\ref{ht}). This effectively sums, in the sense described in Refs.~\cite{parker,jack}, all powers of $R$ and $E(x)$
(with any functional form of coefficients)
in the proper-time series.

When using the above resummed form for the heat-trace, the heat-kernel coefficients become slightly simpler and can be easily written down to relatively high order. The first few are (we are considering the case of a manifold without boundary):
\bea
\mathscr{C}^{(0)}_{\lambda}&=& 1~,\nonumber\\
\mathscr{C}^{(1)}_{\lambda}&=& 0~,\nonumber\\
\mathscr{C}^{(2)}_{\lambda} &=& \mathscr{R} + {1\over 6} \Delta \left(\sigma^2 + \lambda \left|\partial \sigma\right|\right)~,\nonumber
\eea
where 
\bea
\mathscr{R}= {1\over 180}R_{\mu\nu\rho\sigma} R^{\mu\nu\rho\sigma} -{1\over 180}R_{\mu\nu} R^{\mu\nu}- {1\over 120} \Delta R~.\nonumber
\eea
In the above formulas $R$, $R_{\mu\nu}$ and $R_{\mu\nu\lambda\rho}$ are the Ricci scalar and Ricci and Riemann tensors, respectively. Substituting (\ref{ht}) in (\ref{zetan}), one arrives at the following expression for $\zeta(s)$,
\bea
\zeta (s) =  {1\over \Gamma(s)} \sum_{k,\lambda} 
\int_0^\infty dt\, {t^{s-1+k} \over (4\pi t)^{d\over 2}} 
\mathscr{C}_{\lambda}^{(k)} e^{-t\mathscr{X}_{\lambda}} \mathscr{F}_{\beta,\mu}(t)~,
\label{zz}
\eea
where we have defined
\bea
\mathscr{F}_{\beta,\mu}(t) &=& \sum_{n=-\infty}^\infty e^{-t\left( \omega_n^2 -2 i \mu \omega_n-\mu^2\right)}~,\label{ef}\\
\mathscr{X}_{\lambda} &=& \left(R/12 + \sigma^2  + \lambda \left|\partial \sigma\right|\right)~.\label{xx}
\eea
The parameter $s$ works as a regulator and the above expression is
understood as a function in the complex $s$ plane. We assume that $\Re s
< d/2-k -2$ and analytically continue to $s=0$ at the end of the
computation. Under these assumptions the above expression is well defined
and convergent for $\mu=0$.

The dependence
of the effective action from the temperature and the chemical potential
in (\ref{zz}) is factorized. This is a nice bonus of the resummed form
for the heat-trace that we have used and it was noticed before for the
case of zero chemical potential in Ref.\cite{zelnikov}, which also uses a
non-local form for the effective action to analyze finite temperature
effects for free fields in curved space. 

The analytical continuation of $\zeta(s)$ and $\zeta'(s)$ to $s=0$ can be carried out explicitly (the calculation is worked out in Appendix \ref{app1}), leading to the following result:
\bea
\zeta(0)&=&{\beta \over (4\pi)^{D/2}} \sum_{\lambda}
\sum_{k=0}^{[D/2]} \gamma_{k}(D) \mathscr{C}^{(k)}_{\lambda} \mathscr{X}_{\lambda}^{D/2-k}~,
\label{zt0}\\
\zeta'(0) &=& {\beta\over (4\pi)^{D/2}} \sum_{k=0}^\infty \sum_{\lambda}\left(
a_k(D) \mathscr{C}^{(k)}_{\lambda} \mathscr{X}_{\lambda}^{D/2-k}
+\gamma_k(D) \mathscr{C}^{(k)}_{\lambda} \mathscr{X}_{\lambda}^{D/2-k}
\ln \mathscr{X}_{\lambda}\right.\nonumber\\
&&\left.
+
2^{D/2+1-k}
\mathscr{C}^{(k)}_{\lambda}
\left(
{\mathscr{X}_{\lambda}}
\right)^{D/4-k/2} 
\sum_{n=1}^\infty
(-1)^n 
{\cosh(\beta \mu n)
\over
\left(n\beta\right)^{D/2-k}} 
K_{k-D/2}\left(n\beta\sqrt{\mathscr{X}_{\lambda}}\right)
\right)~.
\label{zetaprimeD}
\eea
The coefficients $\gamma_k(D)$ and $a_k(D)$ are given by 
\bea
\gamma_{k}(D) &=& \lim_{s\rightarrow 0}{\Gamma(s+k-D/2)\over \Gamma(s)}~,\nonumber\\
a_k(D)&=& \lim_{s\rightarrow 0}{\Gamma(s+k-D/2)\over \Gamma(s)}\left( \psi^{(0)}\left(s+k-D/2\right) - \psi^{(0)}\left(s\right)\right)~.\nonumber
\eea

It is essential to check the range of convergence of the $n$-summation in (\ref{zetaprimeD}). This can be easily done by expanding the summand for large $n$ as
\bea
(-1)^n {\cosh(\beta \mu n) \over \left(n\beta\right)^{D/2-k}} K_{k-D/2}\left(n\beta\sqrt{\mathscr{X}_{\lambda}}\right) \sim
(-1)^n {e^{n \beta (\mu-\mathscr{X}^{1/2}_{\lambda})} \over n^{1/2}} \left( 1 + O\left({1\over n}\right)\right)~,
\eea
from which we obtain the convergence condition $\Re \left(\mu - \mathscr{X}^{1/2}_{\lambda}\right)<0$. Clearly there is no issue of convergence for purely imaginary $\mu$. However, extending the result to the opposite case, $\Re\left(\mu - \mathscr{X}^{1/2}_{\lambda}\right)>0$, requires some modifications of the approach we used above. We will discuss this in Sec.~\ref{sec5}.

Specializing the result to four dimensions and keeping terms up to second order in the heat-kernel expansion, the results simplify slightly:
\bea
\zeta(0)&=& {\beta\over (2\pi)^{2}} \sum_{\lambda}\left(
{1\over 2}\mathscr{X}_{\lambda}^{2} + \mathscr{C}^{(2)}_{\lambda}
\right)~,\nonumber\\
\zeta'(0) &=& {\beta \over (2\pi)^{2}} \sum_{\lambda}
\left[
{3\over 4} \mathscr{X}_{\lambda}^{2}
-{1\over 2} \mathscr{X}_{\lambda}^{2} \ln \mathscr{X}_{\lambda}
-\mathscr{C}^{(2)}_{\lambda} \ln \mathscr{X}_{\lambda}
\right.
\nonumber\\
&&\left.
+
2\sum_{n=1}^\infty(-1)^n 
\cosh(\beta \mu n)\left(
{4 {\mathscr{X}_{\lambda}} \over \left(n\beta\right)^{2}} 
K_{2}\left(n\beta\sqrt{\mathscr{X}_{\lambda}}\right)
+ \mathscr{C}^{(2)}_{\lambda}K_{0}\left(n\beta\sqrt{\mathscr{X}_{\lambda}}\right)
\right)
\right]~.\nonumber
\eea
In the above expressions, we are neglecting fourth order derivatives of
the condensate, and thus the reminder of the expansion is $O(\partial^{(4)}\sigma)$.

The approach used here presents various advantages. 
The first obvious one lies in the fact
that non-perturbative effects are taken into account.
In flat space, if we use the GL expansion, it
is essential to carry out the calculation at least to fourth
order in the NJL, and sixth order in the GN model
to reproduce the qualitative features of the phase diagram
(Refs.~\cite{nickel1,boehmer} work at sixth order). 
The present approach allows to completely resum various classes of terms in the GL expansion: all terms
proportional to powers of $\sigma$ (and of the
scalar curvature, $R$,) with any functional form of coefficients. 
In this way, non-perturbative effects can be taken into account. 
The advantage to use this resummed approach 
is most stricking when we
consider the case of constant $\sigma$ in flat space. In this case, all
derivatives of $\sigma$ vanish and the result (\ref{zetaprimeD}) becomes
exact ($\mathscr{C}_{\lambda}^{(k)}=0$ for $k\geq 1$). (It is worth
noticing that, although it is not possible to perform a full resummation
of the heat-trace, it is, in fact, possible to engineer the procedure in
a different way and resum the derivatives of the condensate while
expanding in powers $\sigma$.) 

A further bonus lies in the fact that the effective
action turns out to be arranged as the sum of elementary functions plus
a series of $K_\nu(z)$ that decay exponentially. This is rather
advantageous for any numerical manipulation one may have in mind.

Finally, the use of a manifestly covariant regularization allows to
handle renormalization transparently, especially  
when we take into account gravity. 
In contrast to cutoff methods, zeta function regularization 
is manifestly covariant. 
The divergences in the above computation of the determinant 
appear in the form of low order heat-kernel coefficients 
$\mathscr{H}_j$ with $j\leq [D/2]$, 
although the divergences as such are removed in the process of analitic continuation. 
These divergent terms must be tuned 
to experimental inputs. 
In four dimensional space-times, the divergences are proportional to 
$\sigma^2$, $\sigma^4$ and $R\sigma^2$. 
Adjustment of the coefficients of $\sigma^2$ and $\sigma^4$ can 
be achieved by tuning $\ell$ and $G$ in the effective action. 
For the term $R\sigma^2$, we have to introduce an additional term 
$\xi R\sigma^2$ in the effective action 
with a tunable parameter $\xi$. 
As anticipated, experimental input for $\xi$ is 
lacking, and hence it remains unfixed. 
In the example discussed in Sec.~\ref{sec4}, we consider the case of vanishing $\xi$.

We wish to conclude this section by briefly commenting on the strategy we have followed. 
In the previous computation we have performed the sum over $n$ first  and then integrated the
result over $t$. 
Following the opposite order is also possible without any ambiguity due to the
fact that we are in a region of the complex $s$ plane where the expression converges. 
The reader familiar with the work of Chowla and Selberg \cite{chowla} (and some of its generalizations) would have immediately noticed that by integrating expression (\ref{zz}) before performing the summation over $n$ leads to a sum (over $k$) of generalized Epstein-Hurwitz zeta functions. This sum can be recast, by using a generalization of the Chowla-Selberg formula (see Ref.~ \cite{elizalde}), in a form analogous to (\ref{zetaprimeD}). Explicit computation, which we omit here, shows the equivalence of the two results.

\section{Chiral Kinks in static Einstein spaces}
\label{sec4}
The results we have obtained so far are general and allow to obtain the effective action for the condensate.  Any explicit application will require further numerical analysis of the model. This step is non-trivial as it can be understood from the complicated form of the effective action and the goal of this section is to present a sample application to highlight the complexities that may occur in any practical case. In the following, we will consider the case of a four-dimensional static Einstein space, 
\beq
ds^2 = dt^2 -a^2 \left( d\theta^2 + \sin^2\theta \left(d\varphi^2 +\sin^2\varphi d\chi^2\right)\right)~,
\eeq
with $0\leq \theta \leq \pi$, $0\leq \varphi \leq \pi$ and $0\leq \chi \leq 2\pi$. 
Static Einstein spaces have the topology of $\mathbb{R} \otimes \mathbb{S}^{3}$ and constant curvature $R=6a^{-2}$, with $a$ being the radius of the $3$-sphere. 
In the following we will set $\mu=0$ as this simplifies the numerics still permitting to illustrate the technical issues in implementing the formalism.

As mentioned in Introduction, the phenomena of chiral symmetry breaking in $4f$ET in curved space with homogeneous condensates has been considered, for example, in Refs.~\cite{inagaki,eb2,ishikawa,elizalde2,sergeimpla}). In this case the partition function takes a particularly simple form,
\bea
{\mathscr{S}_{eff}\over \beta V} &=& {\sigma^2 \over 2G} - {1\over 2} {1\over (2\pi)^{2}} 
\left[ {3\over 4} \hat{\sigma}^4 -{1\over 2} \hat{\sigma}^4 \ln {\hat{\sigma}^2\over \sigma_0^2} + 8 \hat{\sigma}^2 \sum_{n=1}^\infty (-1)^n 
{\cosh(\beta \mu n) \over \left(n\beta\right)^{2}} K_{2}\left(n\beta \left|\hat{\sigma}\right|\right)
\right]~,\label{zfc}
\eea
with $\hat{\sigma}^2=\left( \sigma^2 +a^{-2}/2\right)$ and $V$ being the volume of $\mathbb{S}^3$. In the above expression we have appropriately rescaled the renormalization length $\ell$ in terms of the constant $\sigma_0$. The flat space limit, $a\rightarrow \infty$ ($\hat\sigma\rightarrow\sigma$), of the above result can be easily compared with the one reported in Refs.~\cite{muta,elizaldebook}, and a simple computation shows the equivalence between these two. 

\FIGURE[t]{
\centerline{
\put(-6,75){$V$}
\put(120,-5){$\sigma$}
\includegraphics[scale=0.8]{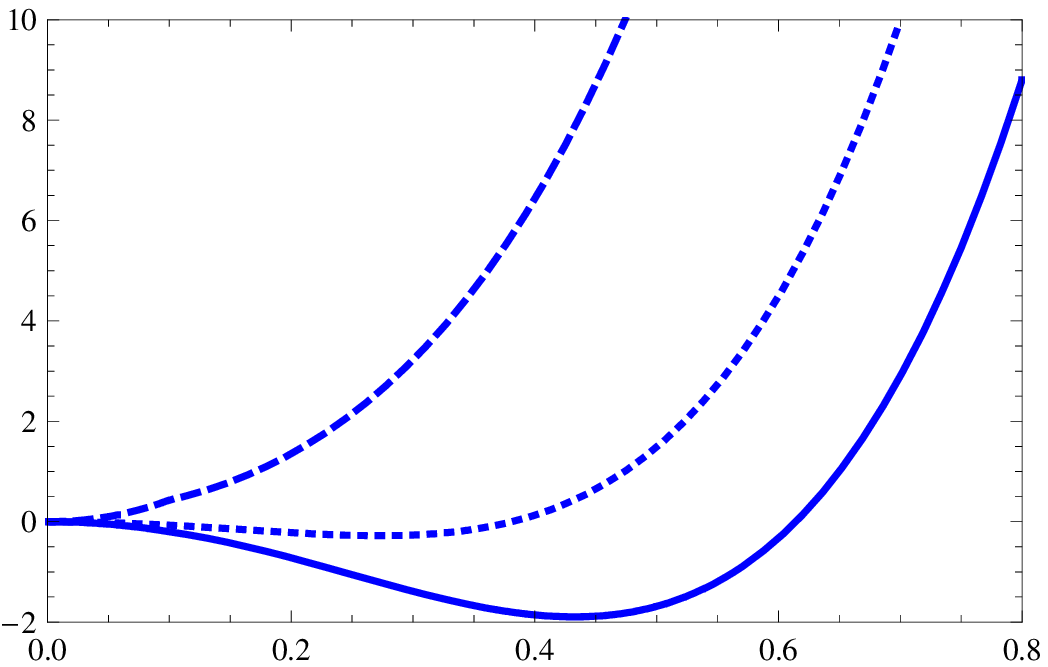}
}
\caption{The figure illustrates the typical behaviour of the thermodynamic potential. The three curves represent the  thermodynamic potential for increasing values of the temperature $T=1$ (continuous line), $T=5$ (short-dashed line), $T=10$ (long-dashed line). The value of the curvature radius is set to $a=10$, the coupling to $G=10$, and the renormalization scale to $\ell=10^3$.}
\label{fig1}
}

Our goal is to include more general cases of inhomogeneous
condensates. We will assume that the condensate $\sigma$ is 
varying only along the $\theta$ direction\footnote{We identify $\gamma_\theta=\gamma_1 a^{-1}$ and this gives $\lambda_\pm= \pm a^{-1}$}, $\sigma\equiv \sigma(\theta)$. A straightforward computation allows one to recast the general result for the effective Lagrangian, including up to second order terms in the heat-kernel expansion, into the following form
\bea
\mathscr{S}_{eff} &=& \int d^Dx \sqrt{g}\left\{ {\tilde\sigma_+^2+\tilde\sigma_-^2\over 4G} - 
{2\over (2\pi)^3}\left[ 
{1\over 2}\left({3\over 2}-\ln \ell^2\right)\left(\tilde\sigma^4_++\tilde\sigma^4_-\right)
-\tilde\sigma^4_+\ln \tilde\sigma_+ 
-\tilde\sigma^4_-\ln \tilde\sigma_-
\right.\right.
\nonumber\\
&&
+ 8 \sum_{n=1}^\infty {(-1)^n \cosh(\beta \mu n) 
\over \left(n\beta\right)^{2}} 
\left(\tilde\sigma^2_+ K_{2}\left(n\beta \left|\tilde\sigma_+\right|\right)
+\tilde\sigma^2_- K_{2}\left(n\beta \left|\tilde\sigma_-\right|\right)\right)
\nonumber\\
&&
\left.\left.+2 \sum_{n=1}^\infty {(-1)^n \cosh(\beta \mu n) \over 6a^2\sin^2\theta} {d^2 \sigma^2\over d\theta^2}
\left( K_{0}\left(n\beta \left|\tilde\sigma_+\right|\right)
+K_{0}\left(n\beta \left|\tilde\sigma_-\right|\right)\right)
\right]\right\}~, 
\label{elesm}
\eea
where $\tilde\sigma^2_\pm=\hat\sigma^2\pm a^{-1}\hat\sigma'$ and the
prime signifies differentiation with respect to $\theta$. The above result can be used as a starting point to analyze the formation of inhomogeneous condensates. 

It is advantageous to notice that under the following rescaling of the parameters,
\bea
\sigma \rightarrow \alpha \sigma~,~~ a \rightarrow \alpha^{-1} a~,~~ \ell \rightarrow \alpha^{-1} \ell~,~~\beta \rightarrow \alpha^{-1} \beta~,~~\mu \rightarrow \alpha \mu~,~~G\rightarrow \alpha^{-2}G~,\nonumber
\eea
the effective action transforms as $\mathscr{S}_{eff} \rightarrow \alpha^4 \mathscr{S}_{eff}$. This allows us to fix one of the quantities aribitrarily without loss of generality.

In the following we will use (\ref{elesm}), the effective Lagrangian to second order in
the heat-kernel expansion. 
Expression (\ref{elesm}) contains sums over $n$ of the form
\bea
\mathscr{A}&=&\sum_{n=1}^\infty {(-1)^n\cosh(\beta\mu n)\over \beta^2n^2} K_2\left(n \beta \left|\tilde\sigma_\pm\right|\right)~,\nonumber\\
\mathscr{B}&=&\sum_{n=1}^\infty {(-1)^n \cosh(\beta\mu n)} K_0\left(n \beta \left|\tilde\sigma_\pm\right|\right)~.\nonumber
\eea
Due to the asymptotic exponential decays of the Bessel functions
$K_\nu(z)$ for large $z$, approximating the sum by its truncated form is 
possible when the arguments are large enough. For inhomogeneous
condensates, however, this approximation does not work uniformly. In fact,
kink-type solutions, that interpolate between two (positive and
negative) extrema cross zero of $\tilde\sigma_\pm$ at around the equator thus invalidating the truncation. This problem can be overcome by replacing the above truncated form 
with a fully resummed expression for $\left|\tilde\sigma_\pm\right|$ smaller than a certain threshold $\varepsilon$. This resummed form can be obtained by expanding the summand assuming $\left|\tilde\sigma_\pm\right|$ is small.
Expanding the Bessel functions to second order in $\left|\tilde\sigma_\pm\right|$, one can obtain the following expressions
\bea
\mathscr{A}&=&{1\over 16}\left[ (\gamma_E-3/4)\left|\tilde\sigma_\pm\right|^4 -\left|\tilde\sigma_\pm\right|^4 \ln \left( 2\beta \left|\tilde\sigma_\pm\right|\right) +{16\over \beta^4} \left( \psi^{(4)}\left(-e^{-\beta \mu}\right) + \psi^{(4)}\left(-e^{\beta \mu}\right)\right)\right.\nonumber \\
&&\left.
-4 {\left|\tilde\sigma_\pm\right|^2\over \beta^2} \left( \psi^{(2)}\left(-e^{-\beta \mu}\right) + \psi^{(2)}\left(-e^{\beta \mu}\right)\right)+{\left|\tilde\sigma_\pm\right|^4} \left( \dot{\psi}^{(0)}\left(-e^{-\beta \mu}\right) + \dot{\psi}^{(0)}\left(-e^{\beta \mu}\right)\right)
\right]
~,\nonumber\\
\mathscr{B}&=&{1\over 2}\left[
\gamma_E -\ln\left(2\beta\left|\tilde\sigma_\pm\right|\right)
+{\beta^2\left|\tilde\sigma_\pm\right|^2\over 4}\left( \dot{\psi}^{(-2)}\left(-e^{-\beta \mu}\right)+\dot{\psi}^{(-2)}\left(-e^{\beta \mu}\right)\right)\right.\nonumber\\
&&\left.+\left( \dot{\psi}^{(0)}\left(-e^{-\beta \mu}\right)+\dot{\psi}^{(0)}\left(-e^{\beta \mu}\right)\right)
\right]~.\nonumber
\eea
In the above expressions we used the notation $\dot\psi^{(a)}(z)=\left.d\psi^{(t)}(z)/dt\right|_{t=a}$. 

In the following we will restrict our analysis to second order in the
derivatives. Then, it is possible to recast the equation of motion for $\sigma$ as a non-linear Schr\"{o}dinger equation (the explicit form for the potential changes depending on whether $\left|\tilde\sigma_\pm\right|>\varepsilon$ or not):
\bea
\sigma'' = \mathscr{U}'\left(\sigma,\sigma'\right)~,
\eea
where the potential $\mathscr{U}\left(\sigma,\sigma'\right)$ depends on $\sigma$, $\sigma'$ and the other parameters of the model. Its explicit expression is rather lengthy and we will not report it here. For homogenous condensates, $\sigma'=0$, the potential has the profile illustrated in Figure~\ref{fig1}. The critical temperature can be computed explictly from the condition that the second derivative of the potential at $\sigma=0$ vanishes. This gives for the critical temperature
\bea
T_{crit}=2\pi\sqrt{3 \pi \over G}~.
\eea
The existence of an inhomogeneous solution depends on the form of the
potential. The same arguments presented by Coleman in the description of
the false vacuum decay are also valid in this case and guarantee the
existence of the solution. However, finding the precise value of the
boundary conditions and explicitly constructing a solution require some
work. Since the effective Lagrangian enjoys $\sigma \leftrightarrow
-\sigma$ symmetry, for kink-type configurations we can search for
solutions in the upper hemisphere,
$\theta\subset\left[\pi/2,\pi\right]$, setting 
$\left.\sigma\right|_{\theta=\pi/2}=0$ and varying
$\left.\sigma'\right|_{\theta=\pi/2}$. The equation of motion for the
condensate is a second order, non-linear differential equation and the value of $\left.\sigma'\right|_{\theta=\pi/2}$
has to be fine-tuned to achieve the correct solution.
Figure~\ref{fig2} illustrates how the solution changes as we increase the
size of the $3$-sphere showing that the kink becomes more steep at the
equator as the curvature increases (The derivativative at the equator is
shown in the right panel of Figure~\ref{fig2}). The behavior of the
solution, {\it i.e.} the fact that it becomes less steep when the radius
$a$ is increased, is mainly due to the fact that $\theta$ is not a
coordinate that measures the proper length. 
If we plot the profile of the kink {\it
vs} the proper distance $a\theta$, 
the profiles obtained for various radii
$a$ almost overlap for fixed values of the temperature. 
For small enough $a$, 
the geometry is no more capable to accomodate a kink. 
This means the existence of a critical value for $a$,
below which kink-type solutions do not exist.  
\FIGURE[t]{
	\centering
\put(-4,65){$\sigma$}
\put(92,-5){$\theta$}
\put(300,-5){$1/a$}
\put(390,65){$\left.{\sigma'\over a}\right|_{\theta={\pi\over 2}}$}
	\includegraphics[width=0.45\columnwidth]{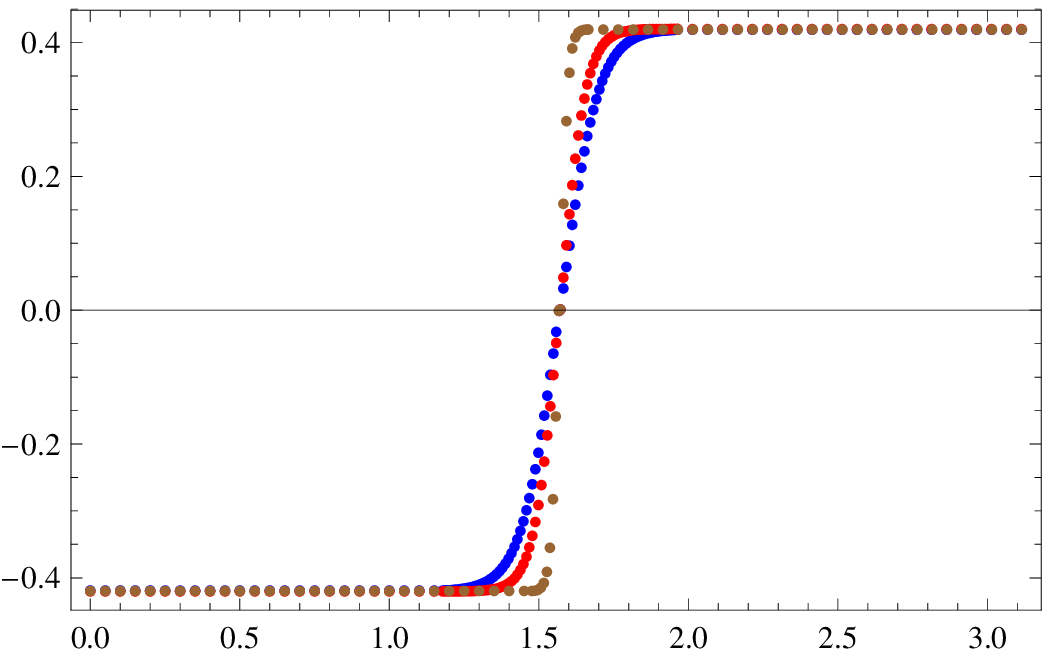}
	\includegraphics[width=0.45\columnwidth]{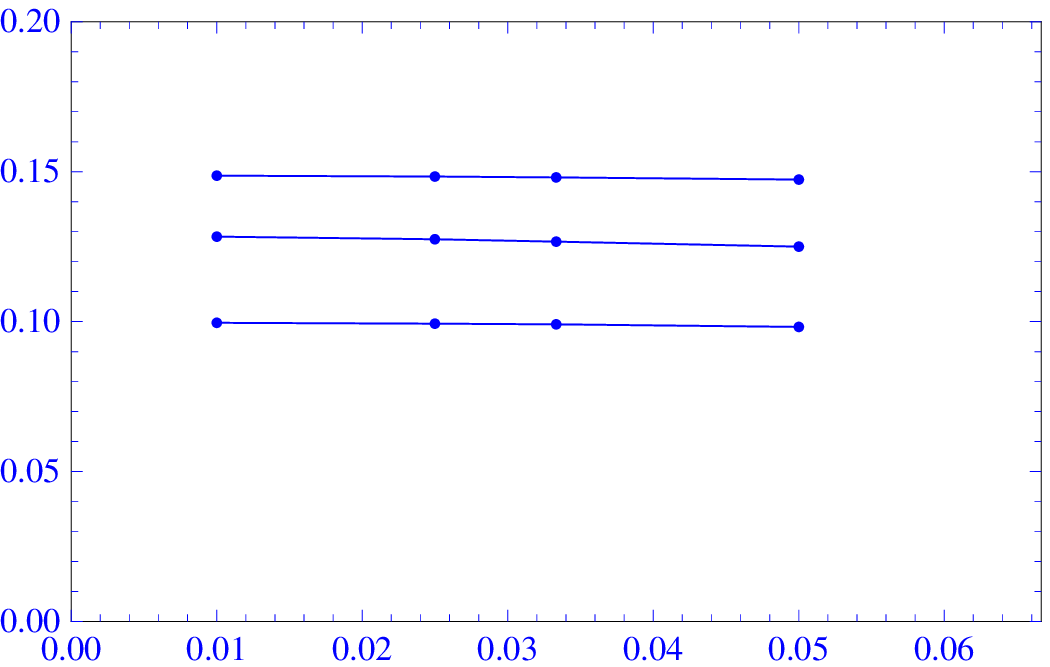}
	\caption{The numerical solution for the kink is shown in the left panel, where the three solutions (right to left) are obtained for $T=2$ and $a=20,~40,~100$. As we increase the radius of the three-sphere (decrease the curvature) the kink becomes more steep and its derivative increase. The behavior of the derivative at the equator is shown in the right pannel for  (top to bottom) 
$T=2,~3,~4$.}
\label{fig2}
}

Aside from kink-type solution, we naturally have homogeneous solutions that correspond to the 
minima of the potential. The phase diagram will be trivial in this case, since the homogeneous solutions are expected to have smaller free energy, thus being favoured compared with inhomogeneous ones. Checking which solution is realized for a smaller/larger value of the action is a simple task. Once the numerical solutions are obtained, the action (in canonical form) 
\bea
S= - \int d\theta \left( f(\sigma) \sigma'^{2} + W(\sigma)\right)~,
\eea
can be computed. Here $f$ and $W$
depend solely on $\sigma$ and not on its derivatives. 
If we shift the
potential in such a way that the minimum corresponding to 
symmetry breaking vacua is at $W=0$, the action
for the homogeneous solution is zero.
Then, the functions $f$ and $W$ are positive definite. 
Therefore inhomogeneous solutions take always larger values
of the action than the homogeneous one,  
indicating that they only appear as meta-stable states.

\section{Chemical Potential}
\label{sec5}
In Sec.~\ref{sec3} we have obtained the effective Lagrangian for a $4f$ET of the form (\ref{action}) at finite temperature and density.
The effective Lagrangian has been obtained in the form of a series
representation, (\ref{zetaprimeD}), which is convergent for
$\Re\left(\mu-\mathscr{X}^{1/2}_{\lambda}\right)<0$. Notice that the representation
(\ref{zetaprimeD}) is always fine for purely imaginary values of
$\mu$. Even inspecting the simple case of homogeneous condensate, the
limitation $\Re\left(\mu-\mathscr{X}^{1/2}_{\lambda}\right)<0$ is clearly not satisfactory
when one needs to consider large densities. The goal of this section is
to illustrate how to adapt the procedure of analytical continuation to
include arbitrarily large, real values of the chemical potential $\mu$ 
in curved space. In fact, this can be achieved in parallel to the procedure in flat space that makes use of the Hubbard-Stratonovich transformation, as done in Ref.~\cite{dgkl}. We note, in passing, that the results of this section also show the equivalence between the series representation of the effective action given in Sec.~\ref{sec3} and the more familiar textbook expression.

The starting point of the present discussion will be the expression (\ref{zzapp1}). 
In our formulation based on zeta functions, we have written the
effective action in terms of $\zeta(0)$ and $\zeta'(0)$. Since
$\zeta(0)$ does not depend on the chemical potential (see formula
(\ref{zt0app1})), it does not require any change. 
As for $\zeta'(0)$, we may write
\bea
\zeta'(0)= {1\over (4\pi)^{d/2}}
      \sum_{k,\lambda} \mathscr{C}^{(k)}_{\lambda}
      {\cal H}^{(k)}_{\lambda}~,
\eea
with 
\bea
{\cal H}^{(k)}_{\lambda}&=&\lim_{s\rightarrow 0}{d\over ds}
  {1\over \Gamma(s)}\int_0^\infty dt\, t^{s-1-d/2+k} e^{-t \mathscr{X}_{\lambda}}
  \mathscr{F}_{\beta,\mu}(t)~.
\label{Hklambda}
\eea
Substituting the explicit form of $\omega_n$ into (\ref{xx}), we have 
\bea
 \mathscr{F}_{\beta,\mu}(t) = 
  {\beta \over \sqrt{\pi t}}
  \sum_{n=-\infty}^\infty (-1)^n e^{-{\beta^2n^2\over 4t}} 
  e^{-\beta \mu n}~.
\nonumber
\eea
Thanks to the presence of the factor $e^{-{\beta^2n^2\over 4t}}$ 
in $\mathscr{F}_{\beta,\mu}(t)$ for $n\ne 0$, 
the integral over $t$ in (\ref{Hklambda}) 
does not produce any additional pole at
$s=0$. Therefore, 
the contribution that remains in the limit $s\to 0$ 
solely comes from the term in 
which $d/ds$ acts on $1/\Gamma(s)\approx s$. 
Thus, we can take the limit $s\to 0$ first for $n\ne 0$. 
Isolating the $n=0$ contribution from the rest of the summation
(signified by the prime in the summation below), we obtain
\bea
{\cal H}^{(k)}_{\lambda}&=&  {\beta \over \sqrt{\pi}}
  \sum_{n=-\infty}^{\infty}\raisebox{5mm}{\!\!\!}' 
   \int_0^\infty dt\,
   {e^{-t \mathscr{X}_{\lambda}}\over t^{(d+3)/2-k} }
   (-1)^n e^{-{\beta^2n^2\over 4t}} e^{-\beta \mu n} 
   +{\cal H}_0~
\eea
where
\bea
{\cal H}_0&=&{\beta \over \sqrt{\pi}}\lim_{s\rightarrow 0}{d\over ds} \int_0^\infty {dt\over \Gamma(s)} t^{s-1-D/2+k} e^{-t \mathscr{X}_{\lambda}}~.
\label{H0}
\eea
Analogously to the case of flat space, 
we may define the density of states, $\rho^{(k)}_{\lambda}$, according to  
\bea
{e^{-t\mathscr{X}_{\lambda}}\over t^{d/2-k}}\equiv 
{1 \over 2}\int_0^\infty dE E \rho^{(k)}_{\lambda} (E) e^{-t E^2}~.
\label{transfdens}
\eea
Using the above expression together with 
the identity
\bea
e^{-tE^2}=2t\int_E^\infty {dx}\, x\, e^{-t x^2}~,
\eea
it is not difficult to arrive at
\bea
{\cal H}^{(k)}_{\lambda}&=& 2
\int_0^\infty dE E\, \rho^{(k)}_{\lambda}\!(E)\, I(E)
   +{\cal H}_0 
\eea
with 
\bea
 I(E) & \equiv & {\beta \over 2\sqrt{\pi}}
  \sum_{n=-\infty}^{\infty}\raisebox{5mm}{\!\!\!}' 
   \int_0^\infty dt\, \int_E^\infty {dx}\, x\, e^{-t x^2} 
   {(-1)^n e^{-{\beta^2n^2\over 4t}} e^{-\beta \mu n}\over \sqrt{t} }
\cr
&=&
  \sum_{n=-\infty}^{\infty}\raisebox{5mm}{\!\!\!}' 
   \int_0^\infty dt\,
     \int_E^\infty {dx}\, x e^{-t x^2}  
    \int dz e^{2\pi i n z} 
    e^{-t\left({2\pi\over \beta}(z+1/2)-i\mu\right)^2}~,
\eea
where we used the Hubbard-Stratonovich transformation to rewrite the
expression for $I(E)$. 
The quantity $I(E)$
can be computed, as in Ref.~\cite{dgkl}. 
Performing the integration over $t$ first, then 
integrating over $z$ using the Cauchy
integral theorem, summing
over $n$, and finally integrating over $x$, 
one easily arrives at the following result 
\bea
I(E) = I_0(E) -
\left[\ln \left({1+e^{-\beta(E-\mu)}\over 1+e^{-\beta E}}
                \right)
 +\ln \left({1+e^{-\beta(E+\mu)}\over 1+e^{-\beta E}}\right)\right]~,
\eea
where $I_0(E)$ corresponds to the expression for $\mu=0$. 
The contribution of $I_0(E)$ combined with ${\cal H}_0$
simply gives the effective action for the $\mu=0$ case, 
which can be calculated as described in Appendix. 
Thus, the final expression for the
effective action becomes
\bea
\mathscr{S}_{eff} &=& 
\left.\mathscr{S}_{eff}\right\vert_{\mu=0}
-{1\over (4\pi)^{d/2}}\int d^d x \sqrt{g} \int_0^\infty dE E \rho(E)
\cr &&
\hspace{3cm} \times\left[\ln \left({1+e^{-\beta(E-\mu)}\over 1+e^{-\beta E}}
                \right)
 +\ln \left({1+e^{-\beta(E+\mu)}\over 1+e^{-\beta E}}\right)\right]~,
\label{Gmu}
\eea
where we have defined 
\bea
\rho(E)= \sum_{k,\lambda} \mathscr{C}_{\lambda}^{(k)}\rho^{(k)}_{\lambda} (E)~.
\eea
As in flat space, the expression (\ref{Gmu}) is regular for any
$\mu$ and can represent the analytic continuation of the effective
action to arbitrary values of the chemical potential on a curved
background. 

To perform the integration over $E$ in Eq.~(\ref{Gmu}), 
we can use 
\bea
\rho^{(0)}_\lambda (E) 
&=& 4 \pi^{-d/2} \int {d^d p}\, \delta\left(E^2 - \mathscr{X}_{\lambda} -p^2\right)~.
\label{densflat}
\eea
By differentiating (\ref{transfdens}) with respect to
$\mathscr{X}_\lambda$, 
we obtain a set of recursive relations:
\bea
\rho^{(k)}_\lambda (E) 
= {\partial^k \over \partial \mathscr{X}_{\lambda}^k} \rho^{(0)}_\lambda
(E)
= (-1)^k{\partial^k \over \partial (E^2)^k} \rho^{(0)}_\lambda (E)~.
\label{513}
\eea
Repeating integration by parts using (\ref{densflat}) and (\ref{513}), 
we can perform the integral over $E$ in Eq.~(\ref{Gmu}), 
leaving $p$-integration. The surface terms that appear from the 
integration by parts can be evaluated by 
using a more explicit expression for $\rho^{(0)}_\lambda (E)$ obtained
by evaluating the $p$-integral in Eq.~(\ref{densflat}):
\bea
\rho^{(0)}_\lambda (E) 
= {2\over \Gamma(d/2)} \left(E^2 - \mathscr{X}_{\lambda}\right)^{d/2-1}~.
\eea
The remaining $p$ integral is manifestly regular
in the $p\to 0$ limit and decays 
exponentially fast for large $p$. Therefore it can be evaluated 
numerically without any difficulties.

\section{Conclusions}
\label{sec6}
Theoretical explorations trying to understand the behavior of hot/dense
strongly interacting matter, as well as experimental attempts to create
a quark-gluon plasma using heavy ion collisions, provide profund
motivations for investigating four-fermion effective theories at finite
temperature and density.

In this context, a great deal of attention has recently been focusing on
the identification of inhomogeneous phases. In fact, although the
subject is not new (see Ref.~\cite{casalbuoni} for a review) and well
explored in condensed matter physics, some very interesting results were
discussed only recently. Two years ago, it was, in fact, demonstrated in
the context of the Gross-Neveu model that, at high densities,
inhomogeneous cristalline phases may form \cite{basar1}. Inhomogeneous
phases have also been recently discussed in the context of Nambu-Jona
Lasinio class of models \cite{nickel1}, where it was suggested that at
high densities the ground state may be populated by a lattice of domain walls implying important changes
in the phase diagram.  Our goal is to study similar phenomena in
curved spacetime, and in this paper we have set the formalism for future
analyses. 

Here, we have followed a less traditional approach and proposed a method
based on the use of the effective action formalism along with zeta
function regularization as basic tools of our analysis. Our approach is
similar in spirit to the work of Ref.~\cite{dgkl} that adopts worldline
Monte Carlo methods to analyze the effective action for strongly
interacting fermionic systems at large $N$ in flat space, particularly
focusing on the Gross-Neveu model. One essential key point of our
approach is the use of a non-pertubative ansatz for the heat-trace that
returns the effective action in a partially resummed form. Within this
approach it is possible to include non-local terms in the effective
action providing a way to go beyond the approximation based on a
truncated form for the Ginzburg-Landau expansion for the partition
function. 
Another important point 
lies in the use of zeta function regularization 
that allows to handle renormalization 
in a manifestly covariant manner, making the regularization procedure 
transparent compared with the cutoff method. 
We have discussed our formalism in the
case of a generic $D$-dimensional ultrastatic Riemannian manifold with
compact spatial section, and obtained an explicit form for the effective
action in terms of a series representation. The series representation
(\ref{las}) is quite advantageous for numerical treatments, since
asymptotically the summands decay exponentially. In fact, the above is
true when the chemical potential is smaller than a certain combination of
the condensate, its derivative, and the curvature. Extending our results
to the general case of real and arbitrarily large chemical potentials
requires some modifications, and to carry out the analytic continuation
for this case, we have adapted the flat space computation of
Ref.~\cite{dgkl}, and worked out a regular expression for the effective
action in this more general case. In passing, we notice that this also
provides a proof of the equivalence between the less standard series
representation of the effective Lagrangian and the standard textbook
formula. 

Implementing our formalism in a specific situation is not immediate. For
this reason, we discussed an application, physically rather simple, but
technically non-trivial, to the case of a four-fermion effective theory
propagating on a static Einstein space at zero chemical potential. 
We have described a way to solve the
effetive equations numerically and, indeed, found kink-type solutions. 
Due to curvature effects we may expect the existence of inhomogenous
phases, but at zero chemical potential 
these kink-type solutions turned out to be energetically less favoured,   
with the free energy larger than that of the corresponding homogeneous 
solutions.

We are currently testing the formalism in various situations. We are
extending the analysis for a static Einstein space with non vanishing
chemical potential. In this case, analogously to flat space case, we
expect a region of the phase diagram where inhomogeneous phases are
energetically favoured. 

While at small curvature the recovery of the flat space behavior is
expected, for large values of the curvature the interplay between
chemical potential and curvature effects becomes non-trivial to
intuit. In addition to that, differently from flat space, 
the finiteness of the spatial section of a static Einstein space 
implies the existence of a critical value of the
curvature beyond which, the geometry cannot accomodate inhomogenous
phases, which will make the structure of the phase diagram richer.
Extending the above results to
anti-de Sitter spacetimes follows straightforwardly. 

Another case we are currently investigating is when an external
potential $V$ is included in the set-up (an interesting case of this
type is a system of strongly interacting fermions in a confining
potential). Adapting the formalism to this case is quite simple and
requires only minor modifications, {\it i.e.} augmenting the functional
determinant in (\ref{las}) by a term of the form $V^2 +
\lambda \left|\partial V\right|$. The heat-kernel coefficients change accordingly
and clearly becomes slightly more involved, but no extra methodological
complication arises. 

Finally, a case that surely deserves attention is that of black
holes. This case is considerably more complex for various reasons. First
of all static black hole spacetimes are only conformally related to ultrastatic
spacetimes. Therefore, in order to include this case, the effective
action requires a correction term (sometimes called cocycle)
\cite{dowker1,dowker2}. A further difficulty arises due to the fact that
black hole spacetimes are not constant curvature ones. This causes
substantial differences even in the case of vanishing chemical
potentials. Technically this does not require any change in the
formalism, but the numerical analysis will be complicated. Additional
problems arise due to the fact that in the present approximation the
effective action diverges at the horizon, analogously to the case
analyzed in Ref.~\cite{page}. This problem can be easily cured by using
a different approximation just in the vicinity of the horizon, but,
again, will make the numerics more involved, since we would need to
match the near-horizon solutions with the outer ones. 
Our analysis is currently in progress and we may anticipate that all
these problems can be solved in the way just mentioned. Details
on the above applications of the formalism presented in this paper will
appear in a follow-up work~\cite{flachitanaka}.

\acknowledgments

We wish to thank Kenji Fukushima for discussions, particularly related to Refs.~\cite{nickel1,nickel2}, and Marco Ruggieri for various explanations of the NJL model and many encouraging conversations. The Japanese Society for Promotion of Science (Grants N. 19GS0219, N. 20740133, and N. 21244033), the Global COE Program ``The Next Generation
of Physics, Spun from Universality and Emergence'', and the Grant-in-Aid for Scientific Research on Innovative Areas (N. 21111006) from the MEXT are gratefully acknowledged for their support.

\appendix

\section{Analytical continuation of $\zeta(s)$ and $\zeta'(s)$}
\label{app1}

This appendix is devoted to carry out explicitly the analytical continuation to $s=0$ of 
\bea
\zeta (s) =  {1\over \Gamma(s)} \sum_{k,\lambda} 
\int_0^\infty dt\, {t^{s-1+k} \over (4\pi t)^{d\over 2}} 
\mathscr{C}_{\lambda}^{(k)} e^{-t\mathscr{X}_{\lambda}} \mathscr{F}_{\beta,\mu}(t)~,
\label{zzapp1}
\eea
where $\mathscr{F}_{\beta,\mu}(t)$ and $\mathscr{X}_{\lambda}$ are defined according to (\ref{ef}) and (\ref{xx}).
We assume that $\Re s < d/2-k -2$ and take the limit $s=0$ at the end. Under these assumptions the above expression is well defined and convergent for any $\mathscr{X}_{\lambda}\geq0$. 

In order to perform the integration over $t$, it is convenient to express the sum over $n$ in terms of the elliptic function $\theta_3$,
\bea
\mathscr{F}_{\beta,\mu}(t) = \sum_{n=-\infty}^\infty e^{-t \left( \omega_n^2 -2 i \mu \omega_n-\mu^2 \right)} &=&  {\beta \over 2\sqrt{\pi t}}\theta_3\left( e^{-\beta^2 \over 4t }; {\pi - i\beta \mu\over 2}\right)~,\nonumber
\eea
and then use the definition
\bea
\theta_3\left(x;y\right) = 1 +2 \sum_{n=1}^\infty x^{n^2} \cos(2ny)~.\nonumber
\eea
The function $\mathscr{F}_{\beta,\mu}(t)$ can then be expressed as
\bea
\mathscr{F}_{\beta,\mu}(t) = {\beta\over 2\sqrt{\pi t}} \left(1 +2 \sum_{n=1}^\infty (-1)^n e^{-\beta^2n^2\over 4t} \cosh(\beta \mu n)\right)~.
\label{fbm}
\eea
Then, we have
\bea
\zeta (s) =  \beta{1\over (4\pi)^{D\over 2}} \sum_{k,\lambda} \left(  \mathscr{C}^{(k)}_{\lambda} \mathbb{I}^{(k)}_{\lambda}(s) 
+ \sum_{n=1}^\infty (-1)^n \cosh(\beta \mu n) \mathscr{C}^{(k)}_{\lambda} \mathbb{J}^{(k,n)}_{\lambda}(s)\right)~, 
\label{zed}
\eea
where
\bea
\mathbb{I}^{(k)}_{\lambda}(s) &=& \int_0^\infty {dt\over \Gamma(s)} t^{s-1-D/2+k} e^{-t \mathscr{X}_{\lambda}}\nonumber\\
&=&
{\Gamma(s+k-D/2)\over \Gamma(s)}\mathscr{X}_{\lambda}^{D/2-k-s}~,\nonumber\\
\mathbb{J}^{(k,n)}_{\lambda}(s) &=& \int_0^\infty {dt\over \Gamma(s)} t^{s-1-D/2+k} e^{-t \mathscr{X}_{\lambda}-{\beta^2n^2\over 4t}}
\nonumber\\
&=& 
{2^{D/2+1-k-s}\over \Gamma(s)} \left({\mathscr{X}_{\lambda}\over n^2\beta^2}\right)^{D/4-(k+s)/2} K_{k+s-D/2}\left(n\beta\sqrt{\mathscr{X}_{\lambda}}
\right)~.\nonumber
\eea
It is easy to check that
\bea
&&\lim_{s\rightarrow 0} \mathbb{I}^{(k)}_{\lambda}(s)= \gamma_k(D) \mathscr{X}_{\lambda}^{D/2-k}~,\nonumber\\
&&\lim_{s\rightarrow 0} \mathbb{J}^{(k,n)}_{\lambda}(s)= 0~,\nonumber
\eea
where
\bea
\gamma_{k}(D)\equiv \lim_{s\rightarrow 0}{\Gamma(s+k-D/2)\over \Gamma(s)}~.
\eea 
With the above results in hands, we find
\bea
\zeta(0)= {\beta\over (4\pi)^{D/2}} \sum_{\lambda}
\sum_{k=0}^{[D/2]} \gamma_{k}(D) \mathscr{C}^{(k)}_{\lambda} \mathscr{X}_{\lambda}^{D/2-k}~.
\label{zt0app1}
\eea
To obtain the above expression, we have used the fact that for $k\geq D/2+1$ the coefficients $\gamma_{k}(D)$ vanish. Also, notice that for odd $D$-dimensional spacetimes, $\gamma_k(D)=0$ for any $k$, thus $\zeta(0)=0$. Keeping only terms up to second order in the heat-kernel expansion and setting $D=4$, we have:
\bea
\zeta(0)= {\beta\over (4\pi)^{2}} \sum_{\lambda}\left(
{1\over 2}\mathscr{X}_{\lambda}^{2} + \mathscr{C}^{(2)}_{\lambda}
\right)~.
\eea
The computation of the derivative of the zeta function is slightly more cubersome, but an explicit expression can be found in a straightforward way using the following relations:
\bea
{d\mathbb{I}^{(k)}_{\lambda}\over ds}(s) &=& -
{\Gamma(s+k-D/2)\over \Gamma(s)}\mathscr{X}_{\lambda}^{D/2-k-s}\left(\ln \mathscr{X}_{\lambda} + \psi^{(0)}(s) + \psi^{(0)}(s+k-D/2) \right)~,\nonumber\\
{d \mathbb{J}^{(k,n)}_{\lambda}\over ds}(s) &=& -
{2^{D/2-k-s}\over \Gamma(s)} \left({\mathscr{X}_{\lambda}\over n^2\beta^2}\right)^{D/4-(k+s)/2} 
\left[ K_{D/2-k-s}\left(n\beta\sqrt{\mathscr{X}_{\lambda}}\right) \ln \left({4\mathscr{X}_{\lambda}\over n^2\beta^2}\right)
\right.\nonumber\\
&&\left.\left.+ 2 K_{k+s-D/2} \left(n\beta\sqrt{\mathscr{X}_{\lambda}}\right)  \psi^{(0)}(s)
+ 2 {d K_\nu\over d\nu}\left(n\beta\sqrt{\mathscr{X}_{\lambda}}\right)\right|_{\nu=D/2-k-s} \right]
~.\nonumber
\eea
In the limit of vanishing $s$, we obtain
\bea
\lim_{s\rightarrow 0} {d\mathbb{I}^{(k)}_{\lambda}\over ds}(s) &=& a_{k}(D) \mathscr{X}_{\lambda}^{D/2-k} + \gamma_k(D) \mathscr{X}_{\lambda}^{D/2-k} \ln \mathscr{X}_{\lambda}~,\nonumber\\
\lim_{s\rightarrow 0} {d\mathbb{J}^{(k,n)}_{\lambda}\over ds}(s) &=& 2^{1-k+D/2}\left({\mathscr{X}_{\lambda}\over n^2\beta^2}\right)^{D/4-k/2} K_{k-D/2}\left(n\beta\sqrt{\mathscr{X}_{\lambda}}\right)~,
\eea
with the coefficients $a_k(D)$ given by
\bea
a_k(D)= \lim_{s\rightarrow 0}{\Gamma(s+k-D/2)\over \Gamma(s)}\left( \psi^{(0)}\left(s+k-D/2\right) - \psi^{(0)}\left(s\right)\right)
~.
\eea
The result for $\zeta'(0)$ is 
\bea
\zeta'(0) &=& \beta{1\over (4\pi)^{2}} \sum_{k=0}^\infty \sum_{\lambda}\left(
a_k(D) \mathscr{C}^{(k)}_{\lambda} \mathscr{X}_{\lambda}^{D/2-k}
+\gamma_k(D) \mathscr{C}^{(k)}_{\lambda} \mathscr{X}_{\lambda}^{D/2-k}
\ln \mathscr{X}_{\lambda}\right.\nonumber\\
&&\left.
+
2^{D/2+1-k}
\mathscr{C}^{(k)}_{\lambda}
\left(
{\mathscr{X}_{\lambda}}
\right)^{D/4-k/2} 
\sum_{n=1}^\infty
(-1)^n 
{\cosh(\beta \mu n)
\over
\left(n\beta\right)^{D/2-k}} 
K_{k-D/2}\left(n\beta\sqrt{\mathscr{X}_{\lambda}}\right)
\right)~.
\label{zetaprimeDapp1}
\eea

\end{document}